\documentclass[twocolumn,floatfix,secnumarabic,amssymb,nobibnotes,aps,prd]{revtex4-2}
\usepackage{amsmath,amssymb}
\usepackage{graphicx}
\usepackage{bm}
\usepackage{color}
\usepackage{mathrsfs}
\usepackage{comment}
\usepackage{float}
\usepackage{tabularx}
\usepackage{ulem}
\usepackage[all]{xy}
\usepackage{tikz-cd}
\usepackage{braket}
\usetikzlibrary{positioning,arrows.meta,decorations.pathreplacing,calc, ,fit,backgrounds}
\newtheorem{theorem}{Theorem}[section]

\newtheorem{definition}[theorem]{Definition}

\newcommand{\R}{\mathbb{R}}
\AtBeginDocument{\colorlet{BLUE}{blue}}

\begin{document}

\title{Reconstructability of Inverse Problems under Symmetry: Separating Structural, Effective, and Physical Upper Bounds}
\author{Isshin Arai}
\email{k078403@kansai-u.ac.jp}
\affiliation{Graduate School of Science and Engineering, Kansai University, Osaka, 564-8680, Japan}
\date{\today}

\begin{abstract}
In our previous work we introduced the \textit{Reconstruction Dimension}, an
upper bound on the reconstructable degrees of freedom determined solely
by the symmetry. 
In actual physical systems, however, the state quantities are generated through physical processes $P_o$ induced by the cause $O$. The state space is therefore often degenerate, and the upper bound determined by the representation-theoretic structure is not reached. This calls for a distinction between the upper bound determined solely
by symmetry and the one realized after the physical process.
We therefore introduce a three-level structure of upper bounds:
the \textit{Structural}, \textit{Effective}, and \textit{Physical
Reconstruction Dimensions} (SRD, ERD, PRD), determined respectively by
the representation-theoretic structure of the intermediate space, by the
concrete design of the reconstruction map, and by the degeneracy of the
state space under the physical process $P_o$.
For each irreducible component they form the hierarchy
$\mathrm{SRD}\ge\mathrm{ERD}\ge\mathrm{PRD}$, so that the two gaps
identify at which level reconstructability is lost.
The framework is illustrated through two contrasting systems: the
orientation dynamics of flake-like particles, where the cause is a velocity gradinet, and the two-body problem, where the cause itself is
defined through the symmetry. This paper provides a concrete framework for reconstruction in physical inverse problems.
\end{abstract}

\maketitle

\section{Introduction}
In physical systems, the problem of estimating or reconstructing the causal relations underlying a system from observed data is generally formulated as an inverse problem~\cite{Eng96, Mac19}.
In this paper, we consider the causal relation
\[
f_O: X \to Y
\]
between state quantities $X$ and $Y$. Here $O$ characterizes the causal relation between the state spaces and is called the \textit{cause}.
We introduce the map
\[
\mathfrak{F}: X \times Y \to O
\]
that reconstructs the cause $O$ from the state pair $(X,Y)$. Hereafter, $\mathfrak{F}$ is called the \textit{reconstruction map}. This class of inverse problems is called \textit{inverse reconstruction}.
Flow visualization that reconstructs flow information from the tracer particles in fluid~\cite{Kru14, Ni15, Cav20, Har24} and  charged-particle $N$-body problem~\cite{Fuc20, Sat21} can be interpreted as inverse reconstructions with symmetry.
Related examples arise in quantum information, such as quantum process tomography~\cite{Poy97, Fuj99, Nie10, Sta24}, as well as in scattering problems~\cite{Co98}.
These problems are not typically described using the present framework, but they can be represented by inverse reconstruction.
%Since the system possesses SO(3) symmetry, SO(3)-equivariance is required of the reconstruction map $\mathfrak{F}$.

In inverse problems, \textit{identifiability}, whether the cause can be uniquely determined from the observables, is often discussed as a central issue~\cite{Mac19, Wi21, Hei25}. In contrast, the inverse reconstruction treated in this paper considers a reconstruction map. This is a standpoint in which the uniqueness of the cause is regarded as resolved or assumed, focusing instead on \textit{reconstructability}: which components of the cause $O$ are reconstructable.

In our previous work~\cite{Arai26G}, we introduced the \textit{Reconstruction Dimension} as an index for evaluating the reachable dimension of a reconstruction map under symmetry. This quantity is determined solely by the representation-theoretic structure under a given map and provides a theoretical upper bound on the reconstructability constrained by symmetry.
This upper bound does not by itself determine the possibility of complete reconstruction or the reconstruction accuracy. In addition, unless the factors constraining reconstructability are well separated, one risks overevaluating reconstructability or misinterpreting the physical meaning of the system. In problems involving equivariant neural networks~\cite{Fuc20, Ge22}%{\color{red}~\cite{Ce21}}
, for instance, such a separation identifies at which stage—representation structure, map design, or physical process—reconstructability is lost.

In this paper, we address this problem by making the process leading to reconstruction explicit and by separating the upper bound on reconstructability into a three-level structure: the Structural Reconstruction Dimension (SRD), the Effective Reconstruction Dimension (ERD), and the Physical Reconstruction Dimension (PRD).
The SRD is equivalent to the quantity derived as the Reconstruction Dimension in our previous work~\cite{Arai26G}, while the ERD and PRD are upper bounds incorporating the constraints due to map design and the physical process, respectively
(definitions are given in Section~\ref{section:dimension}). 

By formulating the reconstruction map $\mathfrak{F}$ as an explicit
object, the upper bound on reconstructability is separated into three
layers, so that one can identify the layer at which reconstructability
is lost.
Whether the loss can be recovered by redesigning the reconstruction map
is determined by which layer it occurs in.
Our main contribution is this hierarchy
$\mathrm{SRD} \ge \mathrm{ERD} \ge \mathrm{PRD}$, obtained by formulating
inverse reconstruction as the XYO+G problem.

The paper proceeds as follows.
Section~\ref{section:inverse-problem} presents the framework of inverse reconstruction (the XYO+G inverse problem) and show three standpoints of physical inverse problems.
Section~\ref{section:dimension} introduces physical processes and observations, and derives the three Reconstruction Dimensions.
Section~\ref{section:example} illustrates the range of the framework through two systems
that differ in symmetry group and in the prior knowledge of the cause:
the orientation dynamics of flake-like particles in fluids and the
two-body problem (worked out for $N = 2$), which take standpoints A and C
of Section~\ref{section:inverse-problem}, respectively.
Section~\ref{section:discussion} organizes three standpoints of inverse problems and situates related approaches within them, and
Section~\ref{section:remark} presents conclusions and outlook.

\section{The XYO+G Inverse Problem}\label{section:inverse-problem}
\begin{definition}[Forward causal problem]
Let
\[
f_o : X \to Y
\]
denote the map, induced by the cause $O$, connecting a state space $X$ to a state space $Y$. This describes a situation in which a state $x \in X$ is projected onto $Y$ by a physical process depending on a value $o \in O$.
\end{definition}

Next, we define \textit{inverse reconstruction} for this causal relation as follows.

\begin{definition}[The XYO+G inverse problem]\label{def:XYO}
The inverse problem in which a "\textit{reconstruction map}" $\mathfrak F$ from the state space $X \times Y$ to the cause $O$,
\[
\mathfrak{F}_{X\times Y, O}: X \times Y \longrightarrow O,
\]
is defined and, in addition, the constraint of a symmetry group $G$ is imposed, is called the XYO+G problem.
\end{definition}

From the viewpoint of group representation theory, 
let $V_X$ and $V_Y$ be the representation spaces of $X$ and $Y$ with respect to the group $G$, and $V_O$ the representation space of $O$ with respect to $G$. Definition~\ref{def:XYO} can then be rewritten as
\[
\mathfrak{F}_{V_X \times V_Y, V_O}: V_X \times V_Y \longrightarrow V_O, \quad +G,
\]
hereafter written simply as $\mathfrak{F}$. Here the direct product $V_X \times V_Y$ is the set of ordered pairs $(v_x, v_y)$,
\[
V_X \times V_Y := \{ (v_x,v_y) \mid v_x \in V_X,\ v_y \in V_Y \},
\]
and subspaces are treated according to an observational constraint.

In the XYO+G problem, defining the group action on $V_X \times V_Y$ for the representation $\rho(g)$ of $g \in G$ by
\[
\rho_{V_X \times V_Y}(g)(v_x, v_y)
=
\bigl(\rho_{V_X}(g)\, v_x,\, \rho_{V_Y}(g)\, v_y\bigr),
\]
the reconstruction map $\mathfrak{F}$ is required to satisfy the following $G$-equivariance condition:
\[
\mathfrak{F} \circ \rho_{V_X \times V_Y}(g) = \rho_{V_O}(g) \circ \mathfrak{F},
\quad \forall g \in G.
\]

The following commutative diagram holds.
\begin{center}
  \includegraphics[width=0.6\linewidth]{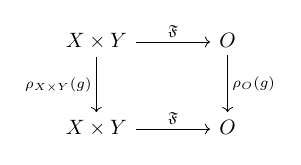}
\end{center}

\begin{definition}[Reconstruction map architecture]
Let $W$ be an intermediate $G$-representation space and let
$\mathcal{H}_G(V_X \times V_Y, W)$ denote the set of nonlinear
$G$-equivariant maps from $V_X \times V_Y$ to $W$. We consider
\[
F \in \mathcal{H}_G(V_X \times V_Y,W),
\]
together with a linear G-equivariant map
\[
F_L \in \mathrm{Hom}_G(W,V_O).
\]
The space $W$ serves as bases extracted from the input state spaces.

In this paper, reconstruction maps are defined as compositions of the form
\[
\mathfrak{F}=F_L\circ F.
\]
The set of all such reconstruction maps is denoted by
\[
\mathfrak{R}_G(V_X\times V_Y,V_O).
\]
\end{definition}

This architecture is not necessarily suitable for every reconstruction map.
Throughout this paper we restrict attention to the set $\mathfrak{R}_G(V_X\times V_Y,V_O)$, in which the linear part $F_L$ is applied after the nonlinear part $F$. By spanning the intermediate space $W$ with bases that carry a clear meaning as vector spaces, and by expressing $V_O$ as linear combinations of their components, the physical interpretation of the reconstruction map becomes explicit. Since $F$ may itself contain linear components or nested compositions of the same type, $F_L\circ F$ can be regarded as the minimal architecture consisting of a nonlinear equivariant part followed by a linear equivariant part.

Hereafter, when such a composite map is treated concretely in the context of design and implementation, we call $F$ a \textit{reconstructor}. A reconstructor can be conceived prior to the detailed structures of its domain and codomain (the state and intermediate spaces).\\

By Schur's lemma~\cite{Ful13},
\[
F_L : W = \bigoplus_{\lambda \in \widehat{G}} W^{(\lambda)} 
\;\longrightarrow\;
V_O = \bigoplus_{\lambda \in \widehat{G}} V_O^{(\lambda)}
\]
is merely a linear map corresponding to the isotypic components. Here $\widehat{G}$ denotes the set of equivalence classes of irreducible representations of $G$ with elements
$\lambda \in \widehat{G}$, and $W^{(\lambda)}$ and
$V_O^{(\lambda)}$ are the isotypic subrepresentations of $W$ and $V_O$, respectively.
That is,
\[
F_L\big|_{W^{(\lambda)}} 
:\; W^{(\lambda)} \longrightarrow V_O^{(\lambda)},
\]
and no mixing occurs between different irreducible components.\\

Forward causal problems appear across many scientific fields, but the prior knowledge about the cause and the information available for estimation or reconstruction differ from field to field.
To clarify the position of this paper, we distinguish three standpoints according to what one aims to obtain:
\begin{description}
\item[A] to obtain the reconstruction map $\mathfrak F$;
\item[B] to obtain the cause $O$;
\item[C] to obtain both $O$ and $\mathfrak F$.
\end{description}
This paper mainly addresses standpoint A. This is the standpoint of seeking an explicit inverse relation (reconstruction map) corresponding to the forward relation among $X, Y, O$. As a result, $O$ is obtained through that relation.
In situations such as the analysis of  flake-like particles  in Section~\ref{subsection:flake}, $\mathfrak F$ can also be created by optimizing the output $O$ against experimentally measured $X, Y$ and $O$.
Standpoint B does not aim at obtaining the relation but at identifying the cause $O$ from the  measured $X$ and $Y$.
Standpoint C is the situation in which one seeks to identify both the cause $O$ and the relation between $O$ and the measured $X$ and $Y$.

Common to these standpoints, the following three problems arise.

\begin{itemize}
\item $Q_1$. Which components of the cause space are identifiable?(\textit{Identifiability}, \textit{Uniqueness})
\item $Q_2$. Which components are reconstructable?(\textit{Reconstructability})
\item $Q_3$. How stable and accurate is the resolution of $O$?
\end{itemize}

In standpoint A, we aims to design the reconstruction map (or reconstructor) $\mathfrak F$ explicitly. When $\mathfrak F$ can be designed, $Q_1$ is satisfied or assumed as a premise. The central problem is thus $Q_2$. In Section~\ref{section:discussion}, we discuss the three standpoints, the problems arising from them.

\section{Structural, Effective, and Physical Upper Bounds in the XYO+G Problem}\label{section:dimension}

%For a map $F \in \mathcal{H}_G(V_X \times V_Y, W)$, let the irreducible decomposition of $W$ be $W = \bigoplus_{\lambda \in \widehat{G}} W^{(\lambda)}$.

\subsection{Structural Reconstruction Dimension}\label{section:limit-cons}

%In the isotropic state space, when the observation space $V_X^C \times V_Y^C$ can completely span each irreducible component of the codomain $W^{(\lambda)}$ of $F$, $\mathrm{Rank}_F$ is
In designing a reconstruction map $\mathfrak{F}\in\mathfrak{R}_G(V_X\times V_Y,V_O)$, the intermediate space $W$---the codomain of $F$ and, at the same time, the domain of $F_L$---plays a central role.
That is, with the projection onto each irreducible component guaranteed by Schur's lemma, the design question is in what sense and in how many copies (bases) the mutually isomorphic irreducible representations should be prepared.
Below we define an upper bound determined solely by the multiplicities of the intermediate space.

Let $m(\lambda)$ denote the multiplicity of $\lambda\in\widehat G$ in $W$, so that
\[
W
=
\bigoplus_\lambda m(\lambda)\, \overline W^{(\lambda)},
\]
where $\overline W^{(\lambda)}$ is an irreducible $G$-representation of type $\lambda$.
This is related to the isotypic components of the previous section by $W^{(\lambda)} = m(\lambda)\,\overline W^{(\lambda)}$. We write $\mathcal{W}$ for the collection of all representation spaces of the group $G$, which is admissible as the intermediate space.

We assume $V_O$ to be multiplicity-free.

\begin{definition}[Card (multiplicity)]\label{def:card}
We define $\mathrm{Card}$ by
\begin{align*}
\mathrm{Card}: \mathcal{W} &\longrightarrow
\bigoplus_{\lambda \in \widehat{G}}
\mathbb{Z}_{\ge 0},\\
W&\longmapsto \Bigl(\min\bigl(m(\lambda),\; \dim \overline W^{(\lambda)}\bigr) \Bigr)_{\lambda \in \widehat{G}}.
\end{align*}
\end{definition}

$\mathrm{Card}$ does not depend on the specific observation spaces of the observation family $\mathfrak{V}^F$ introduced later (Section~\ref{section:state-observe}), nor on concrete designs of maps.

\begin{definition}[Structural Reconstruction Dimension]
To evaluate the upper bound for $W \in \mathcal{W}$ , we define the function
\[
d_W^{(\mathrm{Card})}:\widehat G\longrightarrow \mathbb{Z}_{\ge 0}
\]
by
\[
d_W^{(\mathrm{Card})}(\lambda)
= \mathrm{Card}(W)(\lambda)
\]
for each $\lambda \in \widehat{G}$.
We call $d_W^{(\mathrm{Card})}(\lambda)$ the \textit{Structural Reconstruction Dimension} (SRD)~\cite{Arai26G}.
\end{definition}

From the viewpoint of map design, the SRD represents the upper bound on the dimensions available for spanning $V_O$, determined by the choice of the intermediate space $W$. From the reconstructor viewpoint, it corresponds to the number of output slots into the intermediate space (the number of copies of irreducible representations) at the time the reconstructor is designed.
Since the intermediate space $W$ is determined by the reconstructor, we write $m_F := m$ and $d_F^{(\mathrm{Card})} := d_W^{(\mathrm{Card})}$ when emphasizing this dependence.\\
%Even if one defines $d_{F,P_o}^{(\mathrm{Card})}$ analogously, it is clear that $d_F^{(\mathrm{Card})}=d_{F,P_o}^{(\mathrm{Card})}$.\\

Here,
we consider an example in which each copy of the intermediate space is given meaning through differences in the inputs.
For an input $(v_x, v_y)\in V_X\times V_Y$, an equivariantly designed reconstructor yields an intermediate space with multiplicities $m_F(\lambda)$ as
\begin{equation}
F((v_x, v_y))
=
\bigoplus_{\lambda} F^{(\lambda)}((v_x, v_y)).\label{eq:h}
\end{equation}

Applying this reconstructor in parallel to \(N\) inputs and retaining each output separately defines the reconstructor
\begin{equation*}
\begin{aligned}
F_N((v_x, v_y)_1, & \ldots , (v_x, v_y)_N) \\
&= (F((v_x, v_y)_1), \ldots , F((v_x, v_y)_N))
\end{aligned}
\end{equation*}
The intermediate space then decomposes as
\[
W^{\oplus N}
=
\bigoplus_\lambda N\,m_F(\lambda)\, \overline W^{(\lambda)}.
\]
Hence, the multiplicity of the irreducible representation \(\overline W^{(\lambda)}\) in \(F_N\) is
$N\,m_F(\lambda)$:
\begin{equation}
d_{F_N}^{(\mathrm{Card})}(\lambda)= \min\bigl(N m_F(\lambda),\; \dim \overline W^{(\lambda)}\bigr).\label{eq:eq-Ncard}
\end{equation}
%This represents the analysis performed in Section~\ref{section:tracer-inverse-problem}.

Thus, in a parallel design of a reconstructor,
the SRD increases in proportion to $N$ until it reaches $\dim \overline W^{(\lambda)}$.
In architecture such as the above, where the same reconstructor is used in parallel and different observed states are supplied as inputs, each component of the intermediate space carries the meaning of ``which input it originates from,'' and bases for spanning the cause space $V_O$ are supplied independently for each input; the SRD means an upper bound on the dimensions they can span.

Alternatively, one may vary the design of the reconstructors used in parallel for each input, so that each component of the intermediate space carries the meaning of ``which design it originates from,'' representing the magnitude of the contribution from each design.

\subsection{Physical Processes and Observation}\label{section:state-observe}

A physical process acts on the state space $V_X \times V_Y$. After that, the observation takes place.
We consider physical processes acting on the state space as follows.

\begin{definition}[$P_o$ (physical-process map)]
For each cause $o \in O$, a map
\[
  P_o : V_X \times V_Y \longrightarrow V_X \times V_Y
\]
is given, and the family $\{P_o\}_{o \in O}$ satisfies the equivariance
\[
  P_{\rho_O(g)o}\circ\rho_{V_X \times V_Y}(g)
   = \rho_{V_X \times V_Y}(g)\circ P_o,
  \quad \forall g \in G,\ \forall o \in O.
\]
\begin{center}
  \includegraphics[width=0.8\linewidth]{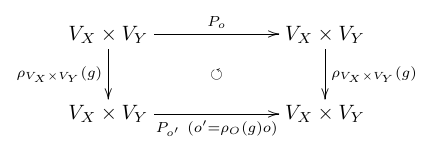}
\end{center}

When $P_o(V_X \times V_Y) \subsetneq V_X \times V_Y$ holds for the isotropic (non-degenerate) state space $V_X \times V_Y$, we call $P_o$ a
physical-process map with degeneracy.
\end{definition}
%Examples of $P_o$ producing pronounced degeneracy include entrainment into attractors and low-energy scattering.

Next, we consider the operation of observing from $V_X \times V_Y$ as a representation of the group $G$.
The observation defined below does not refer to physical observation in the usual sense, such as experimental tools, measurement methods, or measurement errors. Instead, observation in this paper is an abstract concept that represents the selection of a subspace from the input state space based on the reconstruction map.

\begin{definition}[Observation]\label{def:obs-family}
For a nonlinear part $F$,
let $\mathfrak{C}_F$ denote the set of all admissible observations.
Each observation $C \in \mathfrak{C}_F$ determines a finite configuration
\[
V_X^C \times V_Y^C
= \bigl\{ (v_x^i, v_y^i) \bigr\}_{i=1}^{N_C}
\subseteq V_X \times V_Y ,
\]
a finite set of points selected from the state space $V_X \times V_Y$,
where $N_C$ denotes the number of observation pairs.
The family of these configurations,
\[
\mathfrak{V}^F := \bigl\{ V_X^C \times V_Y^C \;\big|\; C \in \mathfrak{C}_F \bigr\},
\]
is introduced, and each $V_X^C \times V_Y^C \in \mathfrak{V}^F$
is called an \textit{observation space}.
We assume $\mathfrak{C}_F$ to be closed under the action of $G$,
i.e. $\rho_{V_X \times V_Y}(g)\bigl(V_X^C \times V_Y^C\bigr) \in \mathfrak{V}^F$
for all $g \in G$.

For each observation $C \in \mathfrak{C}_F$, we define
\[
\widehat{C} : \{V_X \times V_Y\} \longrightarrow \mathfrak{V}^F, \qquad
V_X \times V_Y \longmapsto V_X^C \times V_Y^C .
\]
Note that $\widehat{C}$ does not act on individual points $(v_x, v_y)$.
It is a selection map taking the state space itself as its argument.
\end{definition}

An observation space $V_X^C \times V_Y^C$ corresponds to the finitely many observed data used as inputs to the map, and
$\mathfrak{V}^F$ represents the totality of observation cases that can occur for $F$.
The output of $\widehat{C}$ depends on the contents of the input state space:
if the state space degenerates under a physical process, the corresponding observation space degenerates as well.

The same observation $C$ is applied to the output of $P_o$ after degeneration. It therefore suffices to let $\widehat C$ act on $P_o(V_X \times V_Y)$, and its image defines the observation space $V_X^C \times V_Y^C = \widehat C(P_o(V_X \times V_Y))$. Equivalently, this requires the existence of an induced map $\widetilde P_o$ satisfying $\widehat C\circ P_o
=
\widetilde P_o\circ\widehat C$.

\subsection{Effective and Physical Reconstruction Dimensions}\label{section:limit-bro}

\begin{definition}[$\mathrm{Rank}$ (effective rank)]\label{def:rank}
For a nonlinear part $F$, we define a map that measures, for each irreducible component $\lambda$, the dimension of the subspace of the output representation space $W^{(\lambda)}$ that is accessible from an observation space:

\begin{align*}
\mathrm{Rank} : \mathfrak{V}^F &\longrightarrow
\bigoplus_{\lambda \in \widehat{G}}
\mathbb{Z}_{\ge 0},\\
V_X^C \times V_Y^C &\longmapsto
\Bigl(\dim \operatorname{span} \{F^{(\lambda)}_j(V_X^C \times V_Y^C)\ \big|j = 1,2,..m(\lambda)\}\Bigr)_{\lambda \in \widehat{G}}.
\end{align*}
\end{definition}

The Reconstruction Dimensions based on $\mathrm{Rank}$ are defined as follows.

\begin{definition}[Effective Reconstruction Dimension]\label{def:rank-iden}
For the isotropic state space $V_X \times V_Y$, we define the $\mathrm{Rank}$-based index for an observation $C \in \mathfrak{C}_F$ by
\begin{align*}
\Bigl(\tilde{d}_F^{(\mathrm{Rank} \circ \widehat{C})}(\lambda)\Bigr)_{\lambda \in \widehat{G}}
&:= \mathrm{Rank}\bigl(\widehat{C}(V_X \times V_Y)\bigr)\\
&= \mathrm{Rank}(V_X^C \times V_Y^C)
\in
\bigoplus_{\lambda \in \widehat{G}}
\mathbb{Z}_{\ge 0},
\end{align*}
and, as an upper bound due to map design independent of the observation $C$, we define the function
\[
d_{F}^{(\mathrm{Rank})}:\widehat G \longrightarrow \mathbb{Z}_{\ge 0}
\]
by
\[
d_{F}^{(\mathrm{Rank})}(\lambda)
= \sup_{C \in \mathfrak{C}_F}\,\tilde{d}_F^{(\mathrm{Rank} \circ \widehat{C})}(\lambda)
\]
for each $\lambda \in \widehat G$.
We call $d_{F}^{(\mathrm{Rank})}(\lambda)$ the \textit{Effective Reconstruction Dimension} (ERD).

\end{definition}
Whereas the SRD is an index determined solely by the choice of the intermediate space $W$, the ERD is an index determined by the concrete design of the reconstructor (the actual design of $F$).

\begin{definition}[Physical Reconstruction Dimension]\label{def:rank-phys}
Under the physical-process map $P_o$, we define analogously
\[
d_{F, P_o}^{(\mathrm{Rank})}:\widehat G \longrightarrow \mathbb{Z}_{\ge 0}
\]
by
\[
d_{F, P_o}^{(\mathrm{Rank})}(\lambda)
= \sup_{C \in \mathfrak{C}_F}\,\tilde{d}_F^{(\mathrm{Rank}\circ \widehat{C} \circ P_o)}(\lambda)
\]
for each $\lambda \in \widehat G$. We call $d_{F, P_o}^{(\mathrm{Rank})}(\lambda)$ the \textit{Physical Reconstruction Dimension} (PRD).
\end{definition}

Since the physical-process map with degeneracy $P_o$ breaks the isotropic state of $V_X \times V_Y$ into an anisotropic one,
taking the supremum over observations $C\in\mathfrak{C}_F$ yields a degeneration from the ERD to the PRD.

Together with the SRD, for each $\lambda \in \widehat G$, the hierarchy
\begin{equation}
  d^{(\mathrm{Card})}_{F}(\lambda)
  \;\ge\; d^{(\mathrm{Rank})}_{F}(\lambda)
  \;\ge\; d^{(\mathrm{Rank})}_{F,P_{o}}(\lambda)
  \label{eq:hierarchy}
\end{equation}
holds.
Hereafter we fix $\lambda$ as above, we suppress it from the notation.
The diagram is as follows.
\begin{center}
  \includegraphics[width=0.8\linewidth]{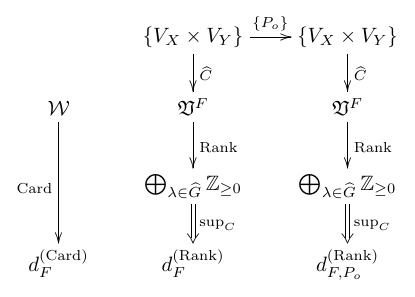}
\end{center}

%\subsection{Relationship among Structural, Effective, and Physical Reconstruction Dimensions}\label{section:card-rank}
Inequality~\eqref{eq:hierarchy} follows from the definitions of Card and Rank. What the two gaps measure, however, does not directly follow from them: each gap isolates a constraint of a different origin through the process leading reconstruction.
The gap
\[
d_F^{(\mathrm{Card})}
-
d_F^{(\mathrm{Rank})}
\]
is the constraint originating from the concrete map design.
In particular, when $F$ fully exploits the structure of the isotropic state space,
the two coincide.
However, as analyzed in the example of Section~\ref{subsection:flake}, blindly increasing the multiplicity of the intermediate space can cause degeneration.
Thus,
whereas the SRD counts the upper bound of the number of copies (multiplicity) $m_F(\lambda)$ in $W$,
the ERD and PRD measure whether those copies actually span linearly independent directions.
The difference between them also stems from redundancy in the design of $F$ (how the tensor products are assembled),
independent of observations and physical processes.
A concrete example of such difference is given in Section~\ref{subsection:flake}.

Furthermore,
PRD is the effective upper bound incorporating the influence of the physical-process map $P_o$, which is necessarily present in physics.
Hence,
\[
d_F^{(\mathrm{Rank})}
-
d_{F,P_o}^{(\mathrm{Rank})}
\]
represents the loss caused by the degeneration of the state space introduced by the physical process.
The equality therefore depends on both $F$ and $P_o$.

A scalar loss cannot distinguish distinct failures: insufficient capacity of the intermediate space, redundancy in the map design, and absence of the information in the state space. The three upper bounds separate them.
Note that these indices take suprema over the observation set $\mathfrak{C}_F$ (ideal observations) and do not represent upper bounds for the actual reconstruction result of each individual observation $C$.
%As described above, inequality~\eqref{eq:hierarchy} is interpreted as a hierarchical structure that separately describes the constraints due to map design and those due to physical processes and observation.
%Although the quantitative evaluation of $d_{F,P}^{(\mathrm{Rank})}$ is generally difficult, the very fact that degeneration occurs carries important meaning in the analysis of inverse problems.

\section{Example}\label{section:example}
In this section we illustrate the framework through two systems that
differ in the prior knowledge of the cause
and in the origin of the degeneracy.
The first (\ref{subsection:flake}), the orientation dynamics of flake-like particles, is treated
from standpoint A, where the cause is a known physical quantity and the
reconstruction map is to be designed.
The second (\ref{subsection:N}), the charged two-body problem, is treated from standpoint C,
where the cause itself is defined through the symmetry.
Together they indicate the range of systems that the XYO+G formulation
covers.

\subsection{Orientation dynamics as an XYO+G}\label{subsection:flake}
Here the cause is assumed to be a known physical quantity. What is to be
designed is the reconstruction map, and the question is which of its
components the design can reach.

Flow visualization using flake-like particles is a classical technique
for qualitative visualization~\cite{Dy82, Got11, Yos23, Itano25},
for which theoretical and experiment]al understandings have been
developed~\cite{Sa85, Gau98}.
Because of hysteresis, however, visualization patterns cannot be directly
associated with instantaneous physical quantities~\cite{Arai24},
which makes the recovery of the governing causal quantity an inverse
problem.
%This system is suited as a first example: the symmetry group acts explicitly, the cause is a known physical quantity, and the state space actually degenerates under the physical process.
In the following, we consider the inverse reconstruction for the physical quantities and introduce the accompanied Reconstruction Dimensions. Specifically, we consider the causal relationship between the orientations of flake-like particles and the velocity gradient tensor that governs their dynamics. Accordingly, we adopt the standpoint of designing a reconstruction map (standpoint A).

Let $\bm{s} \in S^2$ be the unit orientation vector of a flake-like particle.
By the thin-flake approximation of the Jeffery equation for ellipsoidal particles~\cite{Je1922, Br62}, the orientation dynamics obeys
\begin{equation}
  \dot{\bm{s}}
  = J(\bm{s}; A) = \bm{s} \times \bigl(\bm{s} \times (A \cdot \bm{s})\bigr)
  \label{eq:jeffery}
\end{equation}
\cite{Got11}, where $A$ is the local velocity gradient tensor.
The rotational invariance of the cross product gives $J(R\bm{s};RAR^\top)=RJ(\bm{s};A)$,
so the dynamical system is covariant with respect to SO(3). 
Here, we write the $(2\ell+1)$-dimensional irreducible representations of SO(3) as $V_{\ell}$; $R$ is the rotation representation (rotation matrix) acting on $V_1$ (e.g., on the orientation vector $\bm{s}$), acting on Cartesian rank-2 tensors as $A \mapsto RAR^\top$.
The velocity gradient tensor $A$ decomposes under SO(3) action into $V_O=V_0\oplus V_1\oplus V_2$, where $V_0, V_1, V_2$ correspond to the trace (scalar), antisymmetric (vorticity), and symmetric (strain) components, respectively. Therefore, we desgin $\overline W^{(\ell)}$ to carry the representation $V_{\ell}$.\\

The reconstruction map $\mathfrak{F}$ that reconstructs the velocity gradient tensor $A$ from $N$ observation pairs $\{(\bm{s}_i,\dot{\bm{s}}_i)\}_{i=1}^N$ is designed with $F$ consisting of the two layers
\begin{enumerate}
\item[(i)] the bilinear map $\bm{s}_i\otimes\dot{\bm{s}}_i$,
\item[(ii)] the Clebsch--Gordan decomposition $V_1\otimes V_1=V_0\oplus V_1\oplus V_2$,
\end{enumerate}
(Eq.~\eqref{eq:h}), and with
\begin{enumerate}
\item[(iii)] a weighted sum over each irreducible component
\end{enumerate}
as $F_L$.

\begin{figure}[t]
\centering
\includegraphics[width=1\linewidth]{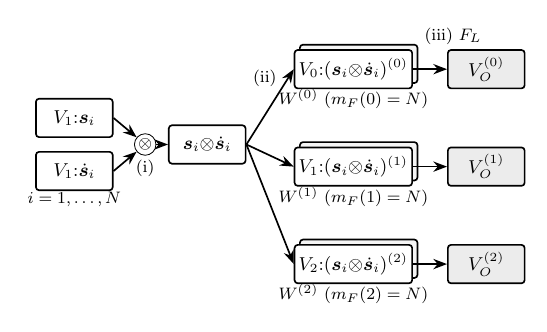}
\caption{%
Reconstruction map for the orientation dynamics, corresponding to layers (i)--(iii) in the text. Stacking $N$ observation pairs gives the intermediate spaces $W^{(\ell)}$ with multiplicity $m_F(\ell)=N$.
}
\label{fig:flake-arch}
\end{figure}

With this three-stage design (Fig.~\ref{fig:flake-arch}), we obtain the parallel architecture described in Section~\ref{section:limit-cons} and in Ref.~\cite{Arai26G}. Ref.~\cite{Arai26G} shows, using an SO(3)-equivariant neural network, a qualitative trend in the reconstruction accuracy of independently isotypic components that is consistent with the SRD.

The multiplicity of each irreducible representation $W^{(\ell)}$ obtained from a single observation pair is $m(\ell)=1$,
and $\dim \overline W^{(\ell)}$ is $2\ell+1$.
The SRD obtained from the parallel design with $N$ observation pairs is, by Eq.~\eqref{eq:eq-Ncard}, $d_F^{(\mathrm{Card})}(\ell) = \min\bigl(N ,\; 2\ell+1\bigr)$ (Table~\ref{tab:card}).
\begin{table}[h]
\centering
\caption{%
  SRD of each irreducible component as a function of the number $N$ of observation pairs.
}
\begin{tabular}{c|ccc}
  \hline
  $N$ & $d_F^{(\mathrm{Card})}(0)$ & $d_F^{(\mathrm{Card})}(1)$ & $d_F^{(\mathrm{Card})}(2)$ \\
  \hline
  1 & 1 & 1 & 1 \\
  2 & 1 & 2 & 2 \\
  3 & 1 & 3 & 3 \\
  4 & 1 & 3 & 4 \\
  5 & 1 & 3 & 5 \\
  \hline
\end{tabular}
\label{tab:card}
\end{table}\\

\begin{figure}[t]
\centering
\includegraphics[width=1\linewidth]{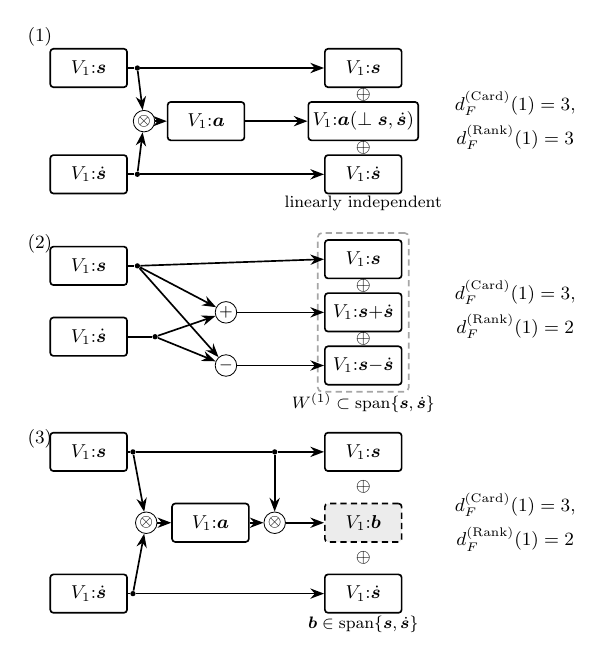}
\caption{%
Three designs of the intermediate space $W^{(1)}$ for a single observation pair $(\bm s,\dot{\bm s})$, corresponding to cases (1)--(3) in the text. Only design (1) attains the SRD; in (2) and (3) the three slots span a two-dimensional subspace. Dashed boxes indicate redundant components.
%Three designs of the intermediate space $W^{(1)}$ for a single observation pair $(\bm s,\dot{\bm s})$, corresponding to cases (1)--(3) in the text. (1) $\{\bm s,\dot{\bm s},\bm a\}$ with $\bm a=\bm s\times\dot{\bm s}$ spans three linearly independent directions, and the ERD attains the SRD. (2) Linear combinations of the inputs, $\{\bm s+\dot{\bm s},\bm s-\dot{\bm s},\bm s\}$, however mutually distinct, remain in the plane spanned by the inputs, so only two independent directions are obtained. (3) Chaining cross products, $\bm b=\bm a\times\bm s=(\bm s\times\dot{\bm s})\times\bm s$, yields a vector lying in the plane spanned by $\bm s$ and $\dot{\bm s}$, so the three slots span only two independent directions. Dashed boxes indicate redundant components.%
}
\label{fig:erd-cases}
\end{figure}

Regarding the ERD, we focus in particular on $\ell = 1$.
As reconstructors, we consider the following three cases (Fig.~\ref{fig:erd-cases}), which differ from the design (i)--(ii) above.
Taking non-parallel input vectors $\bm s,\dot{\bm s}$,
the $V_1$ component of their tensor product is the cross product $\bm a=\bm s\times \dot{\bm s}$.
\begin{enumerate}
\item[(1)] Designing the intermediate space $W^{(1)}$ with $\{\bm s,\dot {\bm s},\bm a\}$, the three directions are linearly independent and fully span $V_1$:
$d_F^{(\mathrm{Card})}(1)=d_F^{(\mathrm{Rank})}(1)=3$.
\item[(2)] Filling the slots with linear combinations of the inputs, e.g. $\{\bm s,\,\bm s+\dot{\bm s},\,\bm s-\dot{\bm s}\}$: although the three slots are mutually distinct, linear operations cannot leave the plane spanned by $\bm s$ and $\dot{\bm s}$, so only two independent directions are obtained:
$d_F^{(\mathrm{Card})}(1)=3,\ d_F^{(\mathrm{Rank})}(1)=2$.
\item[(3)] Chaining tensor products makes the redundancy explicit.
Setting $\bm b=\bm a\times\bm s=(\bm s\times\dot{\bm s})\times\bm s=|\bm s|^2\,\dot{\bm s}-(\bm s\cdot\dot{\bm s})\,\bm s$ shows that $\bm b$ falls back into the plane spanned by $\bm s$ and $\dot{\bm s}$.
Hence, taking $\{\bm s,\bm b,\dot{\bm s}\}$ as the intermediate space, three slots are prepared but only two independent directions are spanned:
$d_F^{(\mathrm{Card})}(1)=3,\ d_F^{(\mathrm{Rank})}(1)=2$.
\end{enumerate}
When a chain of tensor products folds back onto existing directions, $W$ cannot be efficiently spanned.
For the ERD to reach the upper bound given by the SRD,
$F$ must be designed to generate independent directions (appropriate nonlinear/cross terms).
This is a matter of how $W$ is spanned, i.e., of map design.\\

We consider the PRD. In actual physical systems, as a result of time evolution under the Eq.~\eqref{eq:jeffery},
the orientation distribution of particles becomes non-isotropic.
For example, under a steady $A$ the orientations are attracted either to
a great circle or to a fixed point, depending on whether the eigenvalue
of $A$ with the smallest real part is complex or real~\cite{Sze93p,
Sze94, Got11}.
Attractors of the same type have been confirmed under time-varying $A$
by both analysis and numerical simulation~\cite{Arai25, Arai26}.
%In either case the attraction is asymptotically stable, so that after a sufficiently long advection time the orientation distribution is supported on a degenerate subset of $S^2$.
In the former case the orientations and their time derivatives lie in the
same plane, and taking its normal as the polar axis, the $Y_1^0$
component vanishes and $\bm{s}_i,\dot{\bm{s}}_i$ have only $Y^{\pm1}_1$ components in the spherical harmonic basis $Y^m_\ell$.
%Consider the case where all orientation vectors in the flow are restricted to the unit circle $S^1\subset S^2$ and their time derivative vectors are in a same plane.
%In the expansion in the spherical harmonic basis $Y^m_\ell$, the component orthogonal to the plane ($Y^0_1$) vanishes,
The reachable components of each irreducible representation are limited to
\[
  V_0 \ni Y^0_0,\quad
  V_1 \ni Y^{0}_1,\quad
  V_2 \ni Y^{-2}_2 \oplus Y^{0}_2 \oplus Y^{2}_2,
\]
and the $m=\pm1$ components are generated in neither $V_1$ nor $V_2$ for the map design given by (i)--(iii).
Table~\ref{tab:rank} compares the SRD and the PRD.

\begin{table}[h]
\caption{\label{tab:srd-prd-N} SRD and PRD of each irreducible component as functions of the number $N$ of observation pairs, for the design ((i), (ii), (iii)) of the reconstruction map. The PRD is evaluated for an orientation distribution restricted to $S^{1}$. Both bounds saturate in $N$, but at different values.}
\begin{ruledtabular}
\begin{tabular}{c|cccc|cccc}
 & \multicolumn{4}{c|}{SRD} & \multicolumn{4}{c}{PRD} \\
$N\backslash \ell$ & $0$ & $1$ & $2$ & Total & $0$ & $1$ & $2$ & Total \\
\hline
1        & 1 & 1 & 1 & 3 & 1 & 1 & 1 & 3 \\
2        & 1 & 2 & 2 & 5 & 1 & 1 & 2 & 4 \\
3        & 1 & 3 & 3 & 7 & 1 & 1 & 3 & 5 \\
4        & 1 & 3 & 4 & 8 & 1 & 1 & 3 & 5 \\
5        & 1 & 3 & 5 & 9 & 1 & 1 & 3 & 5 \\
$\geq 5$ & 1 & 3 & 5 & 9 & 1 & 1 & 3 & 5 \\
\end{tabular}
\end{ruledtabular}
\label{tab:rank}
\end{table}

Both bounds saturate in $N$, but at different values: the SRD reaches $9$ at $N = 5$, whereas the PRD reaches $5$ already at $N = 3$ and increases no further. Under the $S^1$ restriction the $m = \pm1$ components are absent, so the gap of $4$ persists for arbitrarily many observation pairs.
%Table~\ref{tab:rank} shows that the physical process of restricting the orientation distribution to $S^1$ degrades the Reconstruction Dimension from $9$ as the SRD to $5$ as the PRD. The degeneration of the vorticity component $V_1$ is particularly pronounced (only the $m=0$ direction).
This degeneracy originates from physical processes involving stable fixed-point attractors~\cite{Pi01,Kur03,Guck13}. 
Table~\ref{tab:rank} gives the values for the class of $A$ possessing an
$S^1$ attractor, with hysteresis taken to constitute the entire cause.
To recover the Reconstruction Dimension, one must change the intermediate space or the problem formulation itself so as to prevent the state space from degeneracy, informed by the physical process.

\subsection{Charged two-body problem as an XYO+G}\label{subsection:N}

As another example, we consider the N-body problem,
which has long attracted interest in classical mechanics and is now used as a benchmark for equivariant neural networks
~\cite{Fuc20,Sat21}, and work it out for $N=2$.
Unlike the orientation dynamics of Section~\ref{subsection:flake}, which adopts standpoint A, this example adopts standpoint C.
Here the cause is neither specified nor identified in advance. We design a reconstruction map,
read off the Reconstruction Dimensions, and consider what the resulting
components mean as causes.

Each particle carries a charge $q_i\in\{+1,-1\}$ and evolves under a simple Coulomb interaction.
Let the states of the system at times $0$ and $T$ be
\[
x(0)\in X,\qquad x(T)\in Y,
\]
with $X=Y=(\R^3\times\R^3\times\{\pm1\})^N$ carrying positions, velocities, and charges. The dynamical system is covariant with respect to the Euclidean group E(3).
The time evolution is usually represented as the map
\[
\Phi_T : X \longrightarrow Y,\qquad x(0)\longmapsto x(T),
\]
computed by numerical simulation. More recently, this forward causal problem has also been addressed by graph neural networks and equivariant GNNs.

In the XYO+G formulation, instead of considering the time-evolution map itself, we design a reconstruction map that takes the pair $(x(0),x(T))$ as input.
That is, for the forward relation $f_o : X \to Y$, $+\mathrm{E(3)}$, we consider
\[
\mathfrak F : V_X\times V_Y \longrightarrow V_O,\qquad +\mathrm{E(3)},
\]
where the cause $O$ is interpreted as the features carried by the history of the dynamical evolution in the spirit of standpoint C, $\mathfrak F$ serves as a device that projects the observed pair onto a cause space organized by the symmetry without requiring a pre-identified physical cause.
Taking in particular $\mathfrak F\in\mathfrak R_{\mathrm{E(3)}}(V_X\times V_Y,V_O)$ yields the Reconstruction Dimensions as indices.

Since E(3) contains the translation group on $\mathbb R^3$ and is non-compact, and there are infinite-dimensional irreducible representations, the finite-dimensional framework of this paper does not apply directly.
In practice, however, E(3)-equivariant graph neural networks handle translation equivariance through relative coordinates, and cause (feature) spaces are decomposed by the finite-dimensional irreducible representations of the rotation subgroup O(3).
Moreover, since only positions, velocities, and charges appear here and no parity-distinguishing quantities arise, the irreducible representations of O(3) may be treated as those of SO(3) without essential difference. We write the irreducible representations of SO(3) as $V_{\ell}$ and
decompose the intermediate space of $\mathfrak F$ as $W=\bigoplus_{\ell} m_\ell\,V_{\ell}$.\\

\begin{figure}[t]
\centering
\includegraphics[width=1\linewidth]{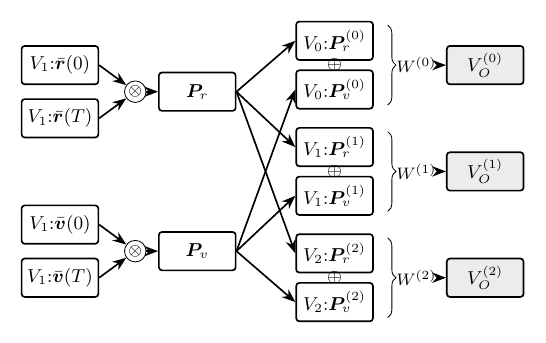}
\caption{%
Reconstruction map for the charged two-body problem. The two-time tensor products $\bm P_r=\bar{\bm r}(0)\otimes\bar{\bm r}(T)$ and $\bm P_v=\bar{\bm v}(0)\otimes\bar{\bm v}(T)$ are decomposed by the Clebsch--Gordan rule, and for each $\ell=0,1,2$ the intermediate space $W^{(\ell)}=\bm P_r^{(\ell)}\oplus\bm P_v^{(\ell)}$ (multiplicity $m(\ell)=2$) is read out linearly to the cause component $V_O^{(\ell)}$.%
}
\label{fig:simple}
\end{figure}

We consider the case $N=2$ with $q_1 q_2=-1$, focusing on a relative motion and thus without translation symmetry.
For the relative position and velocity vectors $\bar{\bm r}$ and $\bar{\bm v}$, the classical solution is the conic
\[
\bar r=\frac{p}{1+e\cos(\theta-\theta_0)},
\]
where $\bar r=|\bar{\bm r}|$ and $\theta$ are polar coordinates in the orbital plane and $\theta_0$ is the orbital phase; the semi-latus rectum $p$ and the eccentricity $e$ are determined by the conserved angular momentum and the energy, with elliptic orbits for $e<1$ and hyperbolic orbits for $e>1$.
The cause $V_O$ is reconstructed from $\bar{\bm r}$ and $\bar{\bm v}$.
As shown in Fig.~\ref{fig:simple}, we design the reconstruction map which forms the two-time tensor products
\[
\bm P_r=\bar{\bm r}(0)\otimes\bar{\bm r}(T),\qquad \bm P_v=\bar{\bm v}(0)\otimes\bar{\bm v}(T),
\]
applies the Clebsch--Gordan decomposition to each channel to obtain
$\{\bm P_r^{(\ell)},\bm P_v^{(\ell)}\}$, $\ell=0,1,2$, and reads out each irreducible component linearly.
Each layer retains the irreducible structure explicitly, and spanning the intermediate space $W$ with bases labeled by the channels gives the cause a physical interpretation.
Since $m(\ell)=2$ for $\ell=0,1,2$, the Reconstruction Dimensions of this simple architecture are as in Table~\ref{tab:N2}.
Note that the multiplicity $m(\ell)=2$ reflects the deliberately simple two-channel design.  Stacking further equivariant tensor products, such as mixed products of $\bar{\bm r}$ and $\bar{\bm v}$ (corresponding to the angular momentum), would increase the multiplicity and hence the SRD, consistent with its character as an upper bound determined by the design policy (Section~\ref{section:limit-cons}).

\begin{table}[!htbp]
\centering
\caption{%
SRD, ERD, and PRD of each irreducible component for the charged two-body reconstruction map of Fig.~\ref{fig:simple} ($m(\ell)=2$).
}
\begin{tabular}{c|ccc}
  \hline
  Component & SRD & ERD & PRD \\
  \hline
  $V_0$  & 1 & 1 & 1 \\
  $V_1$  & 2 & 2 & 1 \\
  $V_2$  & 2 & 2 & 2 \\
  \hline
  Total  & 5 & 5 & 4 \\
  \hline
\end{tabular}
\label{tab:N2}
\end{table}

The degeneration of the PRD to $1$ in $V_1$ reflects angular-momentum conservation in the central-force field. The motion is restricted to a single plane, so that $\bm P_r^{(1)}$ and $\bm P_v^{(1)}$, the antisymmetric components representing the rotational relation between the two times, are aligned with the same angular-momentum direction.
The scalar components $\bm P_r^{(0)}$, $\bm P_v^{(0)}$ carry scalar correlations---distances, phase differences, and velocity correlations---without being in one-to-one correspondence with conserved quantities.
The symmetric traceless components $\bm P_r^{(2)}$, $\bm P_v^{(2)}$ reflect the spatial elongation of the orbit.
Viewed on initial-velocity space, the causal structure in the hyperbolic range is governed primarily by the initial kinetic energy, as reflected by the circular energy contours $|\bar{\bm v}|=\mathrm{const}$.  In the elliptic range, the causal structure is expected to exhibit more intricate dependencies.
Thus the cause space $V_O$ is the decomposition of the dynamical evolution from initial to final state according to the symmetry; each irreducible component provides a distinct information channel, and their direct sum characterizes the transition of the system.

The reconstruction extends hierarchically to larger particle numbers.
Introducing a third particle $(\hat{\bm r},\hat{\bm v},\hat q)$, one forms
\[
\bm Q_r=\hat{\bm r}(0)\otimes\hat{\bm r}(T),\qquad
\bm Q_v=\hat{\bm v}(0)\otimes\hat{\bm v}(T),
\]
and couples them to the two-particle cause space $V_O$ by Clebsch--Gordan products.
For instance, stacking an $\ell=2$ component of the $\bm Q_r$ as the additional particle on the $\bm P_r^{(2)}$ background gives
$V_2\otimes V_2=V_0\oplus V_1\oplus V_2\oplus V_3\oplus V_4$,
generating causal representations up to $\ell=4$.
Increasing the particle number therefore does not redefine the intermediate bases but stacks new representations onto the existing one. The complexity of the cause is described as a hierarchical structure of representation spaces.

Here the cause, not pre-identified, was defined through the reconstruction
map. The
subspaces of $W$ that span it also reflect the conserved quantities, and $W$ may also serve as a reduced space for comparing physical systems.
In this use, however, the irreducibility of its components is not
exploited beyond organizing the degrees of freedom. Such irreducibility may be useful for approximating the forward dynamics such as Map~\eqref{eq:approx} below.

\section{Discussion: Standpoints of Inverse Reconstruction}\label{section:discussion}

In this section, we organize three standpoints.

The mainly problem is then $Q_2$ (Reconstructability) in standpoint A (to obtain the reeconstruction map) such as Section~\ref{subsection:flake} .
When the system possesses symmetry,
the reconstruction map $\mathfrak F$ must be designed with equivariance. Apart from equivariance, there is no particular constraint on the map design, but our map architecture $F_L\circ F$ is an attempt to capture the essence of the system as simply as possible while preserving interpretability in this paper. Within this composition,
decomposing the cause space into irreducible representation spaces allows reconstructability to be analyzed in terms of the Reconstruction Dimension of each irreducible component. The SRD, ERD, and PRD introduced in this paper provide the structural, effective, and physical upper bounds for $Q_2$, respectively.
On the other hand, $Q_3$ is problems of a different level. For example, once the reconstruction map is designed, its stability can be analyzed via the Jacobian and the singular values of the map.\\

Standpoint B is shared by approaches that seek $O$ without constructing
$\mathfrak F$ explicitly. It includes two representative approaches.

The first is statistical estimation, typical of biological systems and medical
inference~\cite{Wi21, Hei25}, where a probabilistic model generating the
observations is assumed. There $Q_1$ (Identifiability) and $Q_3$ are discussed within the frameworks of
likelihood functions, Fisher information, and statistical
identifiability~\cite{Fi1925, Rao45}: $Q_1$ is the non-degeneracy of the Fisher
information, and $Q_3$ is addressed through variance bounds such as the
Cram\'er--Rao inequality. The quantities obtained are determined not by
$\mathfrak F$ but by the assumed distributions and observation model.

The second is optimization- and regression-based estimation, where no
such model is assumed and the cause is defined as the minimizer of a loss function.
As an example, Lagrangian gradient regression
(LGR)~\cite{Har24} estimates the velocity gradient tensor directly from sparse
tracer trajectories. The target is the same as in
Section~\ref{subsection:flake}, whereas the forward relation differs: LGR takes
$(X,Y)$ as neighbouring tracer positions at successive times, while our example takes
$(\bm{s},\dot{\bm{s}})$, so that $f_o$ is the orientation dynamics.
Here $Q_1$ appears as the rank condition that the neighbouring tracer positions
span the physical space, and $Q_3$ is governed by the condition and
the neighbourhood scale. By contrast, $Q_2$ is not explicit. The estimate is a
single tensor, and the reconstructability of individual irreducible components is
not represented.
Quantum process tomography (QPT)~\cite{Nie10, Sta24} occupies both standpoints, and its history illustrates the distinction. The schemes proposed in \cite{Poy97, Fuj99} use the reconstruction map explicitly, which is standpoint A. Contemporary practice instead estimates the process by maximum-likelihood or optimization~\cite{Jev03,Sha11}, where no reconstruction map is written down (standpoint B).\\

Finally, in standpoint C, even the physical meaning of the causal quantity is not pre-identified. In this case, symmetry provides a guideline for organizing the degrees of freedom. Once the symmetry acting on the system is known or assumed, the reconstructable degrees of freedom of the causes can be classified independently by irreducible representation. Using $Q_2$--$Q_3$ as criteria, one can then assess whether each degree of freedom can be meaningful as a causal variable. From the perspective of the classification of unknown causes, this direction is positioned as an extension of inverse reconstruction. It amounts to using $\mathfrak F$ as a device that projects the observables onto an unknown cause space without presupposing the physical nature of the cause. The charged two-body problem of Section~\ref{subsection:N} is an instance of this standpoint. The cause space there is defined not as a set of pre-identified physical quantities but as the symmetry-based decomposition of the dynamical system.

A question that remains open in this standpoint is what evidence supports interpreting the reconstructed quantity as a cause.
In standpoint C the cause space is not
identified with pre-identified physical quantities; the term \textit{cause} is
so far a matter of reconstruction rather than of content.
If the map
\begin{equation}
\tilde{f}_{O} : \widetilde{X} \longrightarrow \widetilde{Y}\label{eq:approx}
\end{equation}
is well-defined on the state spaces $\widetilde{X}$ and $\widetilde{Y}$ constructed from the original state spaces $X$ and $Y$, then the reconstructed quantity $O$ could also be regarded as a cause in terms of its physical content.

\section{Concluding Remarks}
\label{section:remark}
In this paper we formulated inverse reconstruction as the XYO+G problem,
in which the reconstruction map $\mathfrak{F}$ is treated as an explicit
object, and separated the upper bound on reconstructability into the
three levels of the Structural, Effective, and Physical Reconstruction
Dimensions (SRD, ERD, PRD). For each irreducible component these form
the hierarchy
\[
d^{(\mathrm{Card})}_F\ge d^{(\mathrm{Rank})}_F \ge
d^{(\mathrm{Rank})}_{F,P_o} 
\qquad (\text{inequality~\eqref{eq:hierarchy}}),
\]
determined respectively by the choice of the intermediate space $W$, by
the concrete design of the reconstructor, and by the degeneracy of the
state space under the physical process. The two gaps SRD $-$ ERD and
ERD $-$ PRD thus separate the constraint caused by map design
from that caused by the physical process, which makes it possible
to identify at which level reconstructability is lost, and hence how
the loss can be recovered by redesigning the reconstruction map.
These indices provide the directions that
the reconstruction map can reach, and do not guarantee the reconstruction accuracy.
The framework provides guidelines for the formulation of inverse
problems and for the design of reconstruction maps, and can also serve
as a constraint on the search for unknown causes. 

Among
future directions, an important one is the connection to reconstruction or estimation
accuracy. The Reconstruction Dimensions count reachable directions as a
discrete quantity, whereas physical degeneracy is in general continuous. A continuous refinement of these indices
would therefore be expected to connect to existing indices such as the Fisher information. Further directions
include extensions to non-compact groups and to cause spaces that are
not multiplicity-free, and the
classification of unknown causes.

\begin{acknowledgments}
I thank Prof. Tomoaki Itano, Prof. Masako Sugihara-Seki, and Prof. Michihisa Wakui for their valuable guidance and comments on this work.
\end{acknowledgments}

\bibliographystyle{ieeetran}
\bibliography{CardRank}

\begin{thebibliography}{10}
\providecommand{\url}[1]{#1}
\csname url@samestyle\endcsname
\providecommand{\newblock}{\relax}
\providecommand{\bibinfo}[2]{#2}
\providecommand{\BIBentrySTDinterwordspacing}{\spaceskip=0pt\relax}
\providecommand{\BIBentryALTinterwordstretchfactor}{4}
\providecommand{\BIBentryALTinterwordspacing}{\spaceskip=\fontdimen2\font plus
\BIBentryALTinterwordstretchfactor\fontdimen3\font minus
  \fontdimen4\font\relax}
\providecommand{\BIBforeignlanguage}[2]{{%
\expandafter\ifx\csname l@#1\endcsname\relax
\typeout{** WARNING: IEEEtran.bst: No hyphenation pattern has been}%
\typeout{** loaded for the language `#1'. Using the pattern for}%
\typeout{** the default language instead.}%
\else
\language=\csname l@#1\endcsname
\fi
#2}}
\providecommand{\BIBdecl}{\relax}
\BIBdecl

\bibitem{Eng96}
H.~W. Engl, M.~Hanke, and A.~Neubauer, \emph{Regularization of inverse
  problems}.\hskip 1em plus 0.5em minus 0.4em\relax Springer Science \&
  Business Media, 1996, vol. 375.

\bibitem{Mac19}
O.~J. Maclaren and R.~Nicholson, ``What can be estimated? identifiability,
  estimability, causal inference and ill-posed inverse problems,'' \emph{arXiv
  preprint arXiv:1904.02826}, 2019.

\bibitem{Kru14}
D.~Krug, M.~Holzner, B.~L{\"u}thi, M.~Wolf, A.~Tsinober, and W.~Kinzelbach, ``A
  combined scanning ptv/lif technique to simultaneously measure the full
  velocity gradient tensor and the 3d density field,'' \emph{Measurement
  Science and Technology}, vol.~25, no.~6, p. 065301, 2014.

\bibitem{Ni15}
R.~Ni, S.~Kramel, N.~T. Ouellette, and G.~A. Voth, ``Measurements of the
  coupling between the tumbling of rods and the velocity gradient tensor in
  turbulence,'' \emph{Journal of Fluid Mechanics}, vol. 766, pp. 202--225,
  2015.

\bibitem{Cav20}
M.~Cavaiola, S.~Olivieri, and A.~Mazzino, ``The assembly of freely moving rigid
  fibres measures the flow velocity gradient tensor,'' \emph{Journal of Fluid
  Mechanics}, vol. 894, p. A25, 2020.

\bibitem{Har24}
T.~D. Harms, S.~L. Brunton, and B.~J. McKeon, ``Lagrangian gradient regression
  for the detection of coherent structures from sparse trajectory data,''
  \emph{Royal Society Open Science}, vol.~11, no.~10, p. 240586, 2024.

\bibitem{Fuc20}
F.~Fuchs, D.~Worrall, V.~Fischer, and M.~Welling, ``Se (3)-transformers: 3d
  roto-translation equivariant attention networks,'' \emph{Advances in neural
  information processing systems}, vol.~33, pp. 1970--1981, 2020.

\bibitem{Sat21}
V.~G. Satorras, E.~Hoogeboom, and M.~Welling, ``E (n) equivariant graph neural
  networks,'' in \emph{International conference on machine learning}.\hskip 1em
  plus 0.5em minus 0.4em\relax PMLR, 2021, pp. 9323--9332.

\bibitem{Poy97}
J.~Poyatos, J.~I. Cirac, and P.~Zoller, ``Complete characterization of a
  quantum process: the two-bit quantum gate,'' \emph{Physical Review Letters},
  vol.~78, no.~2, p. 390, 1997.

\bibitem{Fuj99}
A.~Fujiwara and P.~Algoet, ``One-to-one parametrization of quantum channels,''
  \emph{Physical Review A}, vol.~59, no.~5, p. 3290, 1999.

\bibitem{Nie10}
M.~A. Nielsen and I.~L. Chuang, \emph{Quantum computation and quantum
  information}.\hskip 1em plus 0.5em minus 0.4em\relax Cambridge university
  press, 2010.

\bibitem{Sta24}
S.~G. Stanchev and N.~V. Vitanov, ``Multipass quantum process tomography,''
  \emph{Scientific Reports}, vol.~14, no.~1, p. 18185, 2024.

\bibitem{Co98}
D.~L. Colton, R.~Kress, and R.~Kress, \emph{Inverse acoustic and
  electromagnetic scattering theory}.\hskip 1em plus 0.5em minus 0.4em\relax
  Springer, 1998, vol.~93.

\bibitem{Wi21}
F.-G. Wieland, A.~L. Hauber, M.~Rosenblatt, C.~T{\"o}nsing, and J.~Timmer, ``On
  structural and practical identifiability,'' \emph{Current Opinion in Systems
  Biology}, vol.~25, pp. 60--69, 2021.

\bibitem{Hei25}
M.~Heinrich, M.~Rosenblatt, F.-G. Wieland, H.~Stigter, and J.~Timmer, ``On
  structural and practical identifiability: Current status and update of
  results,'' \emph{Current Opinion in Systems Biology}, p. 100546, 2025.

\bibitem{Arai26G}
I.~Arai and T.~Itano, ``Group-theoretic upper bounds on reconstructability in
  inverse problems,'' \emph{Physical Review E}, vol. 114, p. 034131, 2026.

\bibitem{Ge22}
M.~Geiger and T.~Smidt, ``e3nn: Euclidean neural networks,'' \emph{arXiv
  preprint arXiv:2207.09453}, 2022.

\bibitem{Ful13}
W.~Fulton and J.~Harris, \emph{Representation theory: a first course}.\hskip
  1em plus 0.5em minus 0.4em\relax Springer Science \& Business Media, 2013.

\bibitem{Dy82}
M.~V. Dyke, \emph{An Album of Fluid Motion.}\hskip 1em plus 0.5em minus
  0.4em\relax Parabolic press, Stanford, California, 1982.

\bibitem{Got11}
S.~Goto, S.~Kida, and S.~Fujiwara, ``Flow visualization using reflective
  flakes,'' \emph{Journal of fluid mechanics}, vol. 683, pp. 417--429, 2011.

\bibitem{Yos23}
K.~Yoshikawa, T.~Itano, and M.~Sugihara-Seki, ``Numerical reproduction of the
  spiral wave visualized experimentally in a wide-gap spherical couette flow,''
  \emph{Physics of Fluids}, vol.~35, no.~3, 2023.

\bibitem{Itano25}
T.~Itano and I.~Arai, ``Probabilistic description of flake orientation
  suspended in rotating wave flows,'' \emph{Physical Review Fluids}, vol.~10,
  no.~9, p. L092901, 2025.

\bibitem{Sa85}
{\"O}.~Sava{\c{s}}, ``On flow visualization using reflective flakes,''
  \emph{Journal of Fluid Mechanics}, vol. 152, pp. 235--248, 1985.

\bibitem{Gau98}
G.~Gauthier, P.~Gondret, and M.~Rabaud, ``Motions of anisotropic particles:
  application to visualization of three-dimensional flows,'' \emph{Physics of
  Fluids}, vol.~10, no.~9, pp. 2147--2154, 1998.

\bibitem{Arai24}
I.~Arai, T.~Itano, and M.~Sugihara-Seki, ``Revisiting visualization of spiral
  states in a wide-gap spherical couette flow,'' \emph{Acta Mechanica}, vol.
  235, no.~12, pp. 7441--7452, 2024.

\bibitem{Je1922}
G.~B. Jeffery, ``The motion of ellipsoidal particles immersed in a viscous
  fluid,'' \emph{Proceedings of the Royal Society of London. Series A,
  Containing papers of a mathematical and physical character}, vol. 102, no.
  715, pp. 161--179, 1922.

\bibitem{Br62}
F.~P. Bretherton, ``The motion of rigid particles in a shear flow at low
  reynolds number,'' \emph{Journal of Fluid Mechanics}, vol.~14, no.~2, pp.
  284--304, 1962.

\bibitem{Sze93p}
A.~J. Szeri, ``Pattern formation in recirculating flows of suspensions of
  orientable particles,'' \emph{Philosophical Transactions of the Royal Society
  of London. Series A: Physical and Engineering Sciences}, vol. 345, no. 1677,
  pp. 477--506, 1993.

\bibitem{Sze94}
A.~J. Szeri and L.~G. Leal, ``Orientation dynamics and stretching of particles
  in unsteady, three-dimensional fluid flows: unsteady attractors,''
  \emph{Chaos, Solitons \& Fractals}, vol.~4, no.~6, pp. 913--927, 1994.

\bibitem{Arai25}
I.~Arai, T.~Itano, and M.~Sugihara-Seki, ``Nonlinear aggregation of phase
  elements on the unit circle under parametric external fields,'' \emph{Journal
  of the Physical Society of Japan}, vol.~94, no.~11, p. 114003, 2025.

\bibitem{Arai26}
I.~Arai and T.~Itano, ``Multifrequency phase locking and hyperplane structures
  in particle orientation,'' \emph{Physical Review E}, vol. 113, no.~1, p.
  014206, 2026.

\bibitem{Pi01}
A.~Pikovsky, M.~Rosenblum, and J.~Kurths, \emph{Synchronization: A universal
  concept in nonlinear sciences}.\hskip 1em plus 0.5em minus 0.4em\relax AIP
  Publishing, 2001, vol.~56, no.~1.

\bibitem{Kur03}
Y.~Kuramoto, \emph{Chemical oscillations, waves, and turbulence}.\hskip 1em
  plus 0.5em minus 0.4em\relax Courier Corporation, 2003.

\bibitem{Guck13}
J.~Guckenheimer and P.~Holmes, \emph{Nonlinear oscillations, dynamical systems,
  and bifurcations of vector fields}.\hskip 1em plus 0.5em minus 0.4em\relax
  Springer Science \& Business Media, 2013, vol.~42.

\bibitem{Fi1925}
R.~A. Fisher, ``Theory of statistical estimation,'' in \emph{Mathematical
  proceedings of the Cambridge philosophical society}, vol.~22, no.~5.\hskip
  1em plus 0.5em minus 0.4em\relax Cambridge University Press, 1925, pp.
  700--725.

\bibitem{Rao45}
C.~R. Rao, ``Information and the accuracy attainable in the estimation of
  statistical parameters,'' \emph{Bull. Calcutta Math. Soc}, vol.~37, no.~3,
  pp. 81--91, 1945.

\bibitem{Jev03}
M.~Je{\v{z}}ek, J.~Fiur{\'a}{\v{s}}ek, and Z.~Hradil, ``Quantum inference of
  states and processes,'' \emph{Physical Review A}, vol.~68, no.~1, p. 012305,
  2003.

\bibitem{Sha11}
A.~Shabani, R.~Kosut, M.~Mohseni, H.~Rabitz, M.~A. Broome, M.~Almeida,
  A.~Fedrizzi, and A.~White, ``Efficient measurement of quantum dynamics via
  compressive sensing,'' \emph{Physical review letters}, vol. 106, no.~10, p.
  100401, 2011.

\end{thebibliography}

\end{document}